\documentclass[letter]{ptptex}
\usepackage{graphics,epsf,bm}
\def\h4l{$^{\hspace*{0.1em}4}_\Lambda$He }
\def\h4s{$^{\hspace*{0.1em}4}_\Sigma$He }
\def\bK{\mbox{\boldmath $K$}}
\def\bR{\mbox{\boldmath $R$}}
\def\bk{\mbox{\boldmath $k$}}
\def\br{\mbox{\boldmath $r$}}
\def\bs{\mbox{\boldmath $s$}}
\begin{document}
\title{Strength of the $\Sigma$ Single-Particle Potential in Nuclei\\
From Semiclassical Distorted Wave Model
Analysis of $(\pi^-,K^+)$ Inclusive Spectrum}
\author{M. Kohno$^1$, Y. Fujiwara$^2$, Y. Watanabe$^3$, K. Ogata$^4$
and M. Kawai$^4$}
\inst{$^1$Physics Division, Kyushu Dental College,
Kitakyushu 803-8580, Japan\\
$^2$Department of Physics, Kyoto University,
Kyoto 606-8502, Japan\\
$^3$Department of Advanced Energy Engineering Science,
Kyushu University, Kasuga, Fukuoka 816-8580, Japan\\
$^4$Department of Physics, Kyushu University,
Fukuoka 812-8581, Japan}
\abst{
Semiclassical distorted wave model is developed to describe
$(\pi^-,K^+)$ inclusive spectra related to $\Sigma^-$-formation
measured at KEK with $p_\pi =1.2$ GeV/c. The shape and magnitude
of the spectrum on $^{28}$Si target are satisfactorily reproduced
by a repulsive $\Sigma$-nucleus potential, the strength of which
is of the order of 30$\sim$50 MeV. This strength is not so strong
as more than 100 MeV suggested by the estimation presented in the
report of the experiment.
}

\maketitle

The study of the $\Sigma$-$N$ interaction is essential
to understanding the physics of octet baryons. In contrast to the
other strange particle $\Lambda$, however, the $\Sigma$
single-particle (s.p.) potential in nuclear medium, which
reflects basic properties of the $\Sigma$-$N$ interaction,
has not been established, since the experimental information has
been scarce. Even the sign of it has been controversial.

$\Sigma$ formation spectra in $(\pi, K)$ or $(K, \pi)$ reactions
on nuclei are not expected to have narrow peaks
because of the strong $\Sigma N\rightarrow \Lambda N$ coupling.
In spite of these circumstances, the early $(\pi, K)$ experimental
spectra \cite{BERT} were interpreted to indicate an attractive
$\Sigma$ s.p. particle potential of around 10 MeV \cite{KHW,DMG}.
The experimental finding of \h4s \cite{HAYA,NAGA} has shown
that the $\Sigma$-$N$ interaction in $T=1/2$ channel is attractive to
support the bound state, as was discussed by Harada \cite{HARA}.
It has been recognized, however, that due to the strong isospin
dependence $\Sigma$ bound states are unlikely to be observed in
heavier nuclei, which has been experimentally supported \cite{BART}.
In the different context,
Batty, Friedman and Gal \cite{BFG} reexamined the $\Sigma^-$ atomic
data to conclude that the $\Sigma$ potential might be repulsive in
a nucleus. The analysis of the $(\pi,K)$ spectra from BNL \cite{BNL}
by D\c{a}browski \cite{DAB} has also suggested that the $\Sigma$
potential is repulsive of the order of 20 MeV.

Theoretical studies have also been elusive for the $\Sigma$-$N$
interaction. In a standard OBEP model for the hyperon-nucleon
interactions, there are uncertainties in the coupling
constants, although $SU_3$ relations are basically imposed.
The Nijmegen group constructed, in 1970's, hard-core
hyperon-nucleon potentials, models D and F \cite{NIJDF}.
Yamamoto and Bando's calculation \cite{YB1} showed that the model D
gave $-16.3$ MeV for the $\Sigma$ potential in nuclear matter
($k_F=1.35$ fm$^{-1}$) and the model F repulsive 5.3 MeV. The later
soft-core version \cite{NIJNS} was shown \cite{YB2} to predict smaller
attraction than the model D.

In recent years, a non-relativistic $SU_6$ quark model has been
developed by Kyoto-Niigata group \cite{FU96a,FU96b,FU01} for the
unifying description of octet baryon-baryon interactions. $G$ matrix
calculations in the lowest order Brueckner theory \cite{KOH}
with this potential showed that the $\Sigma$ s.p. potential
in symmetric nuclear matter is repulsive of the order of 20 MeV
due to the strong repulsion in the $T=\frac{3}{2}$ channel,
which originates from the quark Pauli effects.

Recently, $(\pi^-,K^+)$ spectra corresponding to $\Sigma$ formation
were measured on various nuclei at KEK \cite{KEK} with better
precision, using 1.2 GeV/c $\pi^-$.
The striking report \cite{KEK} in the experimental presentation
of results on $^{28}$Si was that
the $\Sigma$ potential deduced from their DWIA analysis
was strongly repulsive as large as 100 MeV.

The determination of the $\Sigma$-$N$ interaction should have fundamental
influence on such a problem of the neutron star matter and the heavy ion
collision, since the baryonic component of these hadronic mater, especially
the hyperon admixture, is governed by the basic baryon-baryon interactions.
In view of the importance of the understanding of the $\Sigma$-$N$
interaction for our description of the whole octet baryon-baryon interactions,
it is desirable to carry out an independent analysis of the KEK experiments.
In this Letter, we develop a semiclassical method for DWIA approach and apply
it to $(\pi^\pm,K^+)$ inclusive spectra. The semiclassical distorted wave (SCDW)
model was originally considered for describing intermediate-energy nucleon
inelastic reactions on nuclei \cite{SCDW1}. Applications to various
$(p,p')$ and $(p,n)$ inclusive spectra \cite{SCDW2,SCDW3} have proved
that the method works well.

The double differential cross section for the $(\pi ,K)$ hyperon ($Y$)
production inclusive reaction is expressed as
\begin{eqnarray}
\frac{d^2 \sigma}{dW d\Omega} &=& \frac{\omega_i \omega_f}{(2\pi )^2}
 \frac{p_f}{p_i} \int \int d\br d\br ' \sum_{p,h}\;
 \frac{1}{4\omega_i \omega_f}\chi_f^{(-)*}(\br) v_{f,p,i,h}
 \chi_i^{(+)}(\br) \chi_f^{(-)}(\br')\nonumber \\
 &\times& v^*_{f,p,i,h} \chi_i^{(+)*}(\br')
  \phi_p^*(\br) \phi_h (\br) \phi_p(\br') \phi_h^* (\br')
 \delta (W-\epsilon_p + \epsilon_h ) \theta (\epsilon_F -\epsilon_h ).
\end{eqnarray}
where $\chi_i^{(+)}$ and $\chi_f^{(-)}$ describe the incident pion and
final kaon wave functions with energies $\omega_i$ and $\omega_f$,
respectively, and $W=\omega_i -\omega_f$ is the energy transfer.
The $p$ and $h$ denote the unobserved outgoing hyperon ($\Lambda$ or
 $\Sigma$) and nucleon hole states, and the Fermi energy of the
target nucleus is specified by $\epsilon_F$. The transition strength
of the elementary process $\pi + N \rightarrow K + Y$ is represented
by $v_{f,p,i,h}$, which depends on the energy and angle of the
scattering particles, though not explicitly written.
Denoting the c.m. and relative coordinates of $\br$ and $\br'$
by $\bR= \frac{\br + \br'}{2}$ and $\bs= \br'-\br$, respectively,
we introduce a semiclassical approximation:
\begin{eqnarray}
 \chi_f^{(-)}(\bR\pm\frac{1}{2}\bs)
 &\simeq& \mbox{e}^{\pm i\frac{1}{2}\bs\cdot \bk_f (\bR)} \chi_f^{(-)}(\bR),
\\  \chi_i^{(+)}(\bR\pm\frac{1}{2}\bs) &\simeq &
\mbox{e}^{\pm i\frac{1}{2}\bs \cdot \bk_i (\bR)} \chi_i^{(+)}(\bR).
\end{eqnarray}
The hyperon wave functions $\phi_p(\br)$ and $\phi_p(\br')$ are also
treated in the same way.
In these expressions, $\bk(\bR)$ is local classical momentum at the
position $\bR$, which is defined as follows. First, the quantum
mechanical expectation value of the momentum is calculated by
\begin{equation}
 \bk_q(\bR)= \frac{\Re \{\chi^{(\pm)*}(\bR)(-i)\nabla\chi^{(\pm)}(\bR)\}}
 {|\chi^{(\pm)}(\bR)|^2},
\end{equation}
where $\Re$ stands for taking a real part,
and then the magnitude is renormalized by the energy-momentum
relation $\frac{\hbar^2}{2\mu} k^2(\bR) + U_R(\bR) = E$.
$U_R(\bR)$ is the real part of an optical potential which describes the
distorted wave function $\chi$ of each particle with the energy $E$.
Along with the above approximation,
we employ the Thomas-Fermi approximation for the summation of
hole states. Namely, the Bloch density
\begin{equation}
 C(\br,\br';\beta )\equiv\sum_i\phi_i (\br)\phi^*_i (\br') e^{-\beta \epsilon_i},
\end{equation}
where the summation with respect to the single particle state $\phi_i$
of the potential $U$ with the energy $\epsilon_i$ goes over the complete
spectrum, is replaced by that in the Thomas-Fermi approximation \cite{TF}:
\begin{equation}
 C_{TF}(\br,\br';\beta ) = \frac{2}{(2\pi)^3} \int d \bK
 e^{-\beta \{U_\tau (\bR) +\frac{\hbar^2}{2m}\bK^2 \}} e^{i\bK\cdot \bs}.
\end{equation}
The $\theta (\epsilon_F -\epsilon_h)$ function in eq. (1) actually
restricts the integration over $K$ below the local Fermi momentum
 $k_F(\bR)$ defined by the nucleon density $\rho_\tau (\bR)$,
with $\tau$ specifying the proton in the present case, as
$k_F(\bR)=[3\pi^2 \rho_\tau (\bR)]^{1/3}$.

Introducing these approximations, we obtain the following expression
for the double differential cross section, the detailed derivation
of which is discussed in a separate paper:
\begin{eqnarray}
\frac{d^2 \sigma}{dW d\Omega} &=& \frac{\omega_i \omega_f}{(2\pi )^2}
 \frac{p_f}{p_i} \int d\bR \int_{K<k_F(\bR)} d\bK \; \sum_p\;
 \frac{1}{4\omega_i \omega_f} |\chi_f^{(-)}(\bR)|^2 |\chi_i^{(+)}(\bR)|^2 
 |\phi_p (\bR)|^2 \nonumber \\ &\times& |v_{f,p,i,h} (\bK,\bk_i)|^2 
  \delta (\bK +\bk_i(\bR) - \bk_f(\bR)- \bk_p(\bR) )\nonumber \\ &\times&
   \delta (W-\epsilon_p +\frac{\hbar^2}{2m}K^2+U_\tau (\bR)).
\end{eqnarray}
This expression has simple interpretation that the reaction
of $\pi +N$ into $K+Y$ takes place at the position $\bR$
with satisfying the conservation of local semiclassical momenta.
These momenta $\bk_i(\bR)$, $\bk_f(\bR)$ and $\bk_p(\bR)$ are calculated
by eq. (4), using $\pi$, $K$ and $Y$ distorted
wave functions in an optical model description. It should be stressed
that we have avoided the naive introduction of an averaged differential
cross section of the elementary process over the proton momentum
distribution $\rho (\bk )$,
\begin{equation}
\overline{\frac{d\sigma(\pi^- p \rightarrow K^+\Sigma^-)}{d\Omega}}
 \equiv \frac{\int \rho (\bk)
\frac{d\sigma}{d\Omega}(\Omega_k)\delta(k-P) dk}
{\int  \rho (\bk)\delta(k-P) dk}
\end{equation}
with $P=k_K +k_Y -k_\pi$, which was used in the analysis of ref \cite{KEK}.

The transition strength $v_{f,p,i,h}$ is related to the elementary
cross section by
\begin{equation}
 \frac{d\sigma}{d\Omega} = \frac{1}{4\pi} \frac{E_N E_Y}{s}
 \frac{k_K}{k_\pi} |v|^2,
\end{equation}
where $s$ is the invariant mass squared.
We are able to take into account the angular dependence of
the $\pi+ N \rightarrow K +Y$ elementary process.

\begin{figure}[t]
\parbox{\halftext}
{
\epsfxsize=6.6cm
\epsfbox{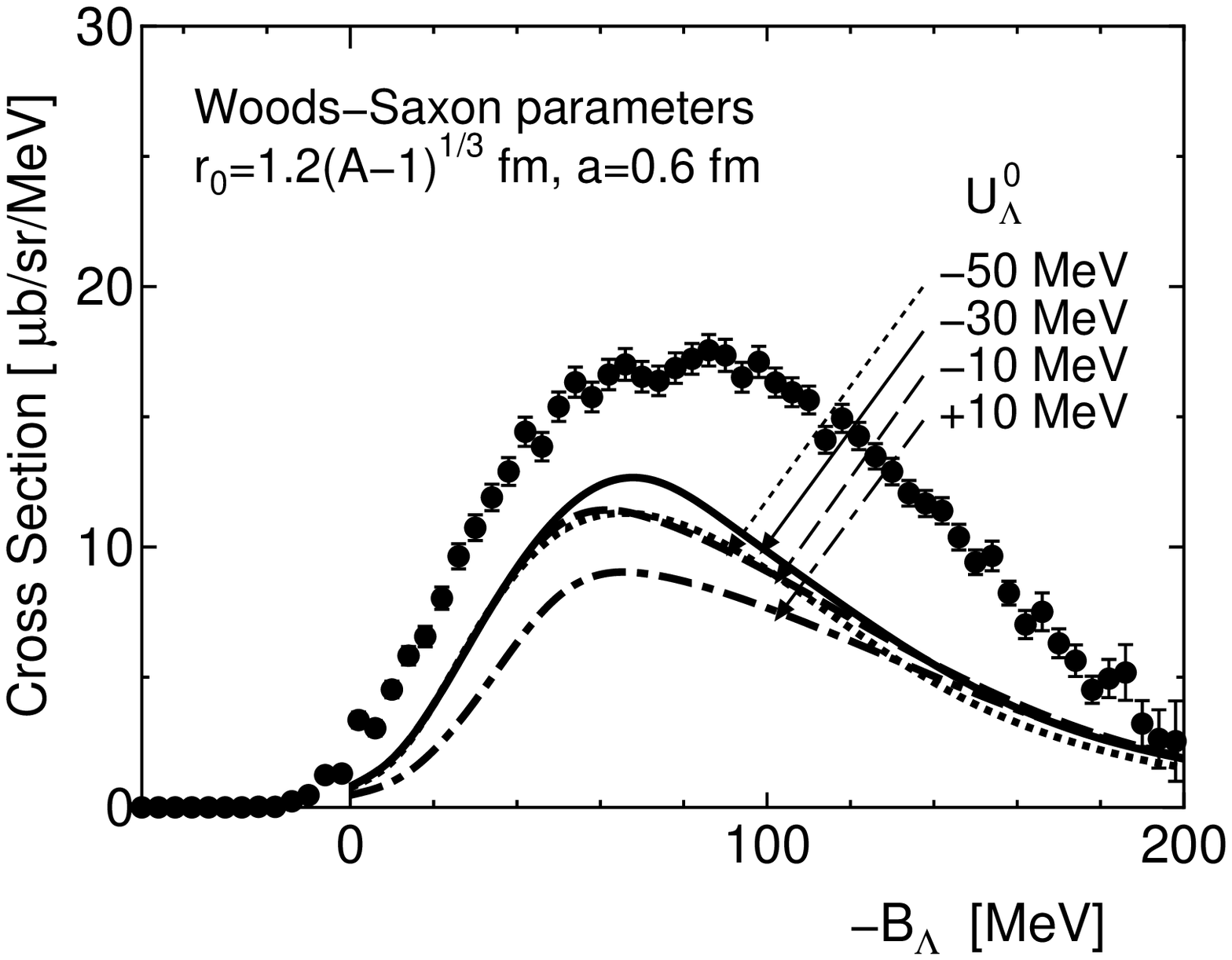}
\caption{$(\pi^+,K^+)$ $\Lambda$ formation inclusive spectra on $^{28}$Si
at $\theta_K =6^\circ \mp 2^\circ$ for the pion of $p_\pi =1.2$ GeV/c,
obtained by various choices of $U_\Lambda^0$ in a Woods-Saxon potential form,
compared with the KEK data \cite{SAHA}.
}}
\hfill
\parbox{\halftext}
{
\epsfxsize=6.6cm
\epsfbox{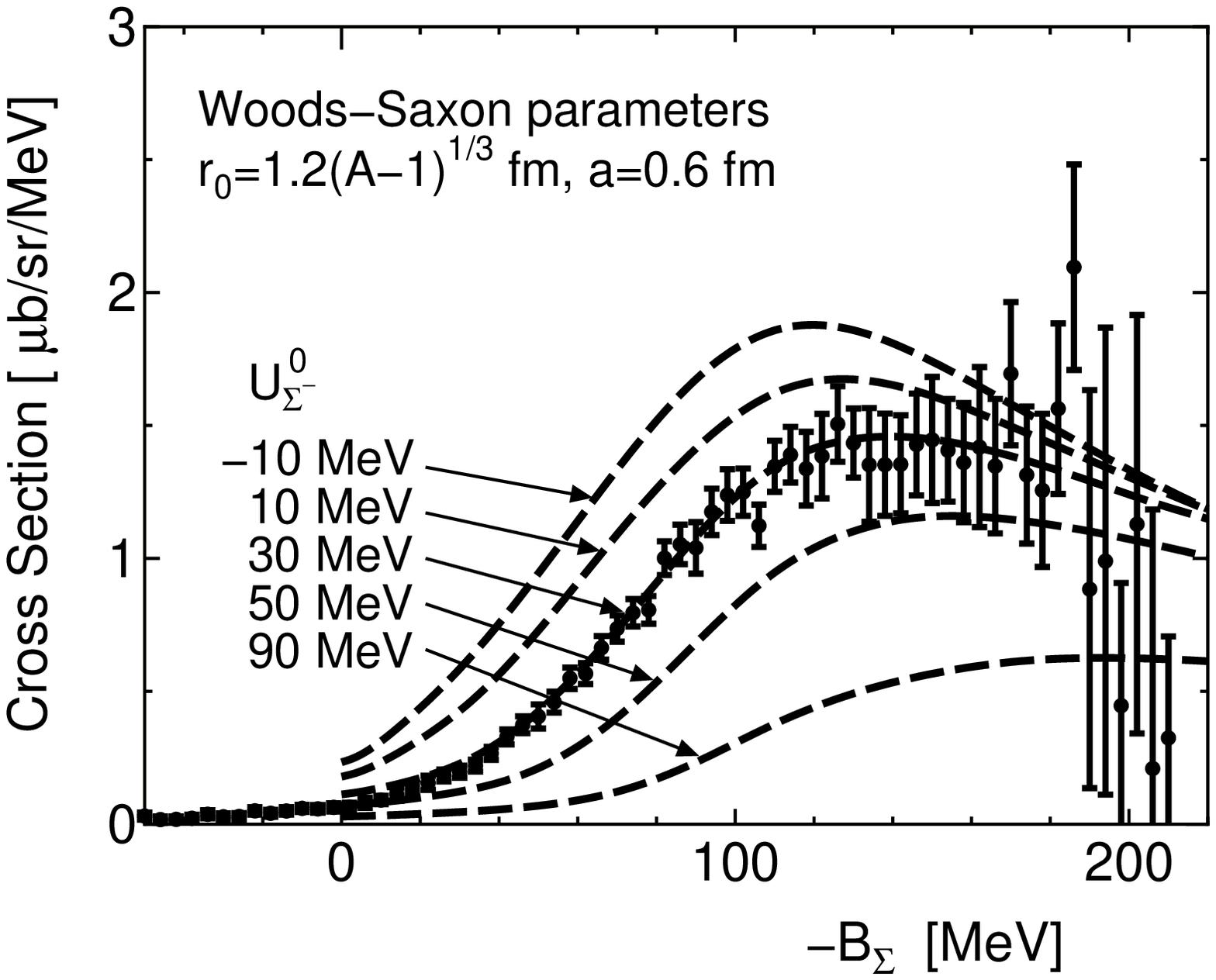}
\caption{$(\pi^-,K^+)$ $\Sigma$ formation inclusive spectra on $^{28}$Si
at $\theta_K =6^\circ \mp 2^\circ$ for the pion of $p_\pi =1.2$ GeV/c,
obtained by various choices of $U_\Sigma^0$ in a Woods-Saxon potential
form, compared with the KEK data \cite{KEK}.
}}
\end{figure}

The local Fermi momentum $k_F(R)$ for $^{28}$Si is prepared with
the nucleon density distribution obtained by the
density-dependent Hartree-Fock method of Campi-Sprung \cite{CS}.
The distorted waves for the incident pion and outgoing kaon are
described by a simple absorptive potential $U(r)=-i \frac{k^2}{2E}
b_0 \rho (r)$ with $\rho (r)$ being the nucleon density distribution.
The parameter $b_0$ is related to the spin-isospin averaged total
cross section of the elementary process by $b_0 \sim \frac{1}{k_\pi}
\langle \sigma_{tot}\rangle$. Referring to the PDG data \cite{PDG}, we
use $b_0=0.58$ fm$^3$ for the incident $1.2$ GeV/c
pion and $b_0=\frac{1}{2p_K} (2.1 (\log p_K -2)+0.84)$ fm$^3$
($p_K$ in fm$^{-1}$) for the outgoing kaon, respectively.

The treatment of the unobserved hyperon is in order.
Actually the hyperon optical potential should be complex,
because inelastic processes are present. Effects of these inelastic
channels may be treated by the Green function method.
Here we adopt simplified prescription to use
a real local potential of the standard Woods-Saxon form
\begin{equation}
 U_Y= \frac{U_Y^0}{1+\exp ((r-r_0)/a)},
\end{equation}
and convolute the result of the calculated spectrum by a Lorentz
type distribution function. The half width is taken to be 5 MeV
for the $\Lambda$ and 20 MeV for the $\Sigma$, based on the imaginary
part of the $\Lambda$ and $\Sigma$ s.p. potentials \cite{KOH}
in symmetric nuclear matter calculated with the quark-model potential FSS.
As the first application, we take the standard geometry parameters:
$r_0=1.2 \times (A-1)^{1/3}$ fm and $a=0.6$ fm. The Coulomb
interaction is incorporated.

We first apply our model to the $(\pi^+,K^+)$ $\Lambda$ formation
inclusive spectrum on $^{28}$Si
measured at KEK \cite{SAHA} with the incident $\pi^+$ momentum
$p_\pi =1.2$ GeV/c. The strength and the angular
dependence of the elementary process are parameterized according to
the available experimental data \cite{LB,SAX}.
Figure 1 shows calculated spectra with various
strengths of the $\Lambda$ s.p. potential
of $V_\Lambda^0 =-50$, $-30$, $-10$ and $10$ MeV, respectively,
to see the potential dependence of the calculated spectra,
though the $\Lambda$ s.p. potential has been established as
$V_\Lambda^0 \sim -30$ MeV from various $\Lambda$ hypernucear data.
Bearing in mind various ambiguities in the elementary amplitudes
which would be modified in nuclear medium and additional two-step
contributions, our model is seen to be capable to describe the
inclusive spectra.

Figure 2 shows calculated $\Sigma^-$ formation $(\pi^-, K^+)$
inclusive spectra, compared with the KEK experimental data.
In the present calculation, we assume an isotropic
angular dependence. The energy dependence
of $|v|^2$ is taken from the parameterization by Tsushima {\it et al.}
\cite{THF}. Their overall strength was normalized by factor 0.82
to match the experimental data taken at KEK \cite{KEK}.
Several curves correspond to the supposed $\Sigma$ potential
of $U_\Sigma^0=-10$, $10$, $30$, $50$ and $90$ MeV, respectively.
No overall renormalization factor is introduced. It is
seen that the shape and absolute value are satisfactorily reproduced
by the repulsive strength of 30 MeV. Expecting the contributions from
multi-step processes, the repulsive strength may be larger to be 50 MeV,
though it is premature to draw
the final conclusion before taking into account various effects
discussed later. This order of the repulsive magnitude of the $\Sigma$
s.p. potential is in line with the estimation by D\c{a}browski \cite{DAB}
for BNL data that the $\Sigma^-$-nucleus potential for $^9$Be is
about 20 MeV. It was noted in ref. \cite{KEK}
that the peak position at as high as 150 MeV was hard to
reproduce if the repulsion of the $\Sigma$-nucleus potential is not
so strong. Present calculations suggest, however, that the
$\Sigma$ s.p. potential is not necessary to be strongly repulsive.
The attractive potential fails by predicting much
larger cross sections. On the other hand stronger repulsive $\Sigma$
s.p. potential tends to underestimate the cross section.
The cause of the different result from ref. \cite{KEK}
might be due to avoiding the use of the factorization approximation
by the average cross section, eq. (18). This point deserves further
investigation.

As for the $SU_6$ quark model,
it is interesting to observe that this order of magnitude of a few ten MeV
is consistent with the prediction of the calculation in nuclear matter
\cite{KOH}. The quark model description of the $\Sigma N$ interaction predicts
definite strong repulsive nature in the isospin $T=\frac{3}{2}$ channel
due to the quark antisymmetrization effect. Thus the $\Sigma^-$-nucleus
potential is expected to become more repulsive in the circumstance
of the neutron excess. In this respect, the analysis of the $(\pi^-,K^+)$ data
on heavier nuclei would be interesting to investigate whether such
quantitative isospin dependence actually exists.

There are various simplified treatments in the present
calculations. The smearing by the Lorentz type convolution should
be replaced by the Green's function method. Quantitative estimation
of the contribution from multistep processes is needed.
More sophisticated description of the elementary process
has to be employed. Contributions from more than two-step
processes tend to increase the cross section. On the other hand,
the possible modification of the elementary process
in nuclear medium would reduce the cross section. These problems
are future subjects to be investigated.
The present SCDW framework serves as a quantitatively reliable
model to discuss
possible change of properties of intermediate baryonic states.

In summary, we have developed a semiclassical distorted wave model
for $(\pi,K)$ inclusive spectra corresponding to the $\Lambda$
or $\Sigma$ formation.
The expression of the double differential cross section consists of
the incoming pion distorted wave function, outgoing kaon distorted
wave function and undetected hyperon distorted wave function
at each collision point, where the conservation of the classical
local momentum is respected. The bound nucleon in the target nucleus is
described by a local Fermi gas model.
The present framework is easily applied to describe other inclusive
spectra, such as $(K, \pi)$, $(\pi, \eta)$, $(\gamma, K)$ and so on.
It is also straightforward to consider multistep contributions, as was
carried out for $(p,p')$ and $(p,n)$ inclusive spectra \cite{SCDW2,SCDW3}.
The application of this model to the $(\pi^- ,K^+)$ inclusive
spectrum on $^{28}$Si taken at KEK \cite{KEK} has shown
that the spectrum is satisfactorily reproduced by a repulsive
$\Sigma$-nucleus potential of the order of 30$\sim$50 MeV. This magnitude,
in turn, constrains the $\Sigma$-$N$ potential model to improve our
understanding of the interactions between complete octet baryons.

\bigskip

The authors would like to thank H. Noumi and T. Harada for valuable
discussions. This study is supported by Grant-in-Aids for Scientific
Research (C) from the Japan Society for the Promotion of
Science (Grants No. 15540284 and No. 15540270).

\end{document}